\documentclass[usenatbib]{mnras}

\usepackage{newtxtext,newtxmath}
\usepackage[T1]{fontenc}
\usepackage{ae,aecompl}

\usepackage{graphicx,graphics}	
\usepackage{amsmath,amssymb}	
\usepackage{color}
\usepackage{url}
\usepackage{hyperref}

\usepackage[T1]{fontenc}

\def\aj{{AJ}}                   
\def\araa{{ARA\&A}}          
\def\apj{{ApJ}}                 
\def\apjl{{ApJ}}                
\def\apjs{ {ApJS}}               
\def\ao{ {Appl.~Opt.}}           
             
\def\aap{ {A\&A}}

\def\mnras{ {MNRAS}}

\newcommand{\swift}{\textit{Swift}}

\newcommand{\xmm}{\textit{XMM-Newton}}

\newcommand{\be}{\begin{equation}}
\newcommand{\ee}{\end{equation}}

\newcommand{\gtsima}{$\; \buildrel > \over \sim \;$}
\newcommand{\ltsima}{$\; \buildrel < \over \sim \;$}
\newcommand{\prosima}{$\; \buildrel \propto \over \sim \;$}
\newcommand{\gsim}{\lower.5ex\hbox{\gtsima}}
\newcommand{\lsim}{\lower.5ex\hbox{\ltsima}}
\newcommand{\simgt}{\lower.5ex\hbox{\gtsima}}
\newcommand{\simlt}{\lower.5ex\hbox{\ltsima}}
\newcommand{\simpr}{\lower.5ex\hbox{\prosima}}

\newcommand{\cds}{$\cdots$}

\title[UV/X-ray properties of Coma UDGs]{Ultraviolet and X-ray Properties of Coma's Ultra-Diffuse Galaxies}
\author[Lee, Hodges-Kluck \& Gallo]{Chris H. Lee$^{1,3}$\thanks{E-mail: chl2019@rit.edu}, 
Edmund Hodges-Kluck$^{2}$,
Elena Gallo$^{3}$\\
$^1$ Center for Imaging Science, Rochester Institute of Technology, Rochester, NY 14623, USA\\
$^2$ Code 662, NASA GSFC, Greenbelt, MD 20771, USA\\
$^3$ Department of Astronomy, University of Michigan, 1085 S University, Ann Arbor, MI 48109, USA\\
}

\begin{document}
\maketitle

\begin{abstract}
Many ultra-diffuse galaxies (UDGs) have been discovered in the Coma cluster, and there is evidence that some, notably Dragonfly 44, have Milky Way-like dynamical masses despite dwarf-like stellar masses. We used X-ray, UV, and optical data to investigate the star formation and nuclear activity in the Coma UDGs, and we obtained deep UV and X-ray data (\textit{Swift} and \textit{XMM-Newton}) for Dragonfly 44 to search for low-level star formation, hot circumgalactic gas, and the integrated emission from X-ray binaries. Among the Coma UDGs, we find UV luminosities consistent with quiescence but NUV$-r$ colors indicating star formation in the past Gyr. This indicates that the UDGs were recently quenched. The $r$-band luminosity declines with projected distance from the Coma core.
The Dragonfly 44 UV luminosity is also consistent with quiescence, with SFR$<6\times 10^{-4} M_{\odot}$~yr$^{-1}$, and no X-rays are detected down to a sensitivity of $10^{38}$~erg~s$^{-1}$. This rules out a hot corona with a $M > 10^8 M_{\odot}$ within the virial radius, which would normally be expected for a dynamically massive galaxy.
The absence of bright, low mass X-ray binaries is consistent with the expectation from the galaxy total stellar mass, but it is unlikely if most low-mass X-ray binaries form in globular clusters, as Dragonfly 44 has a very large population. Based on the UV and X-ray analysis, the Coma UDGs are consistent with quenched dwarf galaxies, although we cannot rule out a dynamically massive population.
\end{abstract}


\section{Introduction}

Hundreds of ultra-diffuse galaxies (UDGs) have been recently discovered in the Coma cluster. They are typically spheroidal in shape with central surface brightness $\mu_{g,0} = 24-26$~mag~arcsec$^{-2}$ and have large effective radii (of a few kpc) for their stellar masses \citep{vdokkum2015, koda2015}. Most UDGs adhere to the low mass end of the red sequence, indicating that they are passively evolving cluster members \citep{yagi2016, zaritsky2019}. UDGs are remarkable for their potentially high dark matter fractions ($\gtrsim98\%$), whose large gravitational potentials may protect the diffuse stellar component from tidal disruptions by the cluster environment \citep{vdokkum2016, koda2015}.

Although UDGs were first reported by \citet{binggeli1985} in the Virgo Cluster, their low surface brightness and low concentration makes them difficult to detect, and it is only recently that large populations have been discovered. These larger studies began with the Dragonfly (DF) Telephoto Array by \cite{vdokkum2015}, which found 47 UDGs in the Coma Cluster. This DF database constitutes the sample that is investigated in this paper, but subsequently more UDGs have been discovered in Coma. Prompted by the DF study, \cite{koda2015} found 854 UDGs in the $\sim4.1^{\circ}\times4.1^{\circ}$ field of view (FOV) of archival Subaru Prime Focus Camera data. Since the DF FOV is about twice the size, one expects about 1,000 UDGs in its field, provided equally sensitive data. \cite{yagi2016} extended the analysis with the Subaru data to specify the UDG parameters. \cite{zaritsky2019} further expanded the search area using the DECam Legacy Survey fields in a 10$^{\circ}$ radius around Coma, finding about 300 UDGs when using a more conservative threshold for the effective radius, $r_{\text{eff}} \gtrsim 2.5$~kpc. This sample constitutes the SMUDGes survey (Systematically Measuring Ultra-Diffuse Galaxies). Beyond Coma, UDGs have been identified in other galaxy clusters and in the field. 

The Coma UDGs are indeed different from other galaxies. \cite{vdokkum2017} found that Coma UDGs have $\sim$7 times more globular clusters (GCs) than galaxies of same luminosity, while \cite{danieli2018} found that the S\'ersic index of faint large galaxies decreases with magnitude (opposite to that of bright, large galaxies). Meanwhile, the stellar populations appear to be old and metal poor \citep{gu2018}, with little to no star formation activity \citep{singh2019}.

One Coma UDG, DF44, has received particular attention because its large number of GCs and stellar velocity dispersion imply a Milky Way-like dynamical mass of $10^{12} M_{\odot}$, whereas its stellar mass of $3\times 10^8 M_{\odot}$ is consistent with a dwarf galaxy \citep{vdokkum2015,vdokkum2016,vdokkum2017,gu2018}. This implies an extreme dark matter fraction exceeding 99\%, which challenges traditional galaxy models, especially as DF44 may be representative of larger UDGs. 

The origin of UDGs remains unclear. \cite{jiang2019} conducted smoothed-particle hydrodynamic simulations of both field and cluster UDGs, concluding that cluster UDGs become quiescent through ram-pressure stripping of cold gas by the intracluster medium and become diffuse through tidal distortions near pericenter. They find that the star-formation rate (SFR) and $B-R$ color gradient depend on the distance to the cluster center. In contrast, they find that field UDGs may occur through gas depletion by supernova feedback, which predicts that only galaxies of a certain mass range become UDGs in the field. Meanwhile, \cite{liao2019} used the Auriga cosmological magneto-hydrodynamical simulations to investigate the formation of UDGs of Milky Way-sized galaxies. They studied the morphology and visual properties of the UDGs, such as sizes, central surface brightnesses, S\'ersic indicies, colors, spatial distribution, and abundance. Their simulations suggest that field UDGs form in ``high-spin'' dark matter halos and that supernova feedback is less important, but that half of UDGs become diffuse through tidal interactions.

Ultraviolet (UV) and X-ray observations can provide crucial clues to UDG formation. From the perspective of galaxy assembly, the UV band is sensitive to even low levels of star formation, while X-rays trace X-ray binaries (which scale with the stellar population) and can show whether UDGs exist in hot halos of gas, as expected for relatively massive galaxies \citep{white&frenk1991}. X-rays can also identify even weakly accreting supermassive black holes, which have yet to be discovered in UDGs and whose absence in Milky Way-sized galaxies would be surprising. \cite{singh2019} recently studied the UV properties of SMUDGes UDGs using GALEX data, finding few detections and concluding that there is no star formation. Meanwhile, \cite{kovacs2019} reported X-ray non-detections for isolated UDGs using \textit{XMM-Newton} survey fields that cover Subaru survey fields \citep{greco2018}. 

Here we examine the UV and X-ray properties of the Coma DF survey galaxies  \citep{vdokkum2015} with GALEX and the \textit{Chandra} X-ray Observatory. We restrict ourselves to this sample because the archival UV and X-ray coverage and sensitivity is highest within $\sim$1~Mpc of the cluster core. We also report on new observations of DF44 with the \textit{Neil Gehrels Swift Observatory} (\textit{Swift}) and \textit{XMM-Newton} (XMM). These deep data enable us to sensitively search for ongoing or recent star formation, X-ray binaries, hot gas, and an AGN. DF44 is presently the best Coma UDG target for X-ray observations because of its potentially large dynamical mass (predicting a hot halo) and its large projected distance from the Coma cluster center (1.8~Mpc from the Coma core in projection and separated by 900~km~s$^{-1}$ from the systemic velocity), which reduces the background from the hot intracluster medium and makes it more likely that the galaxy can hold onto its hot gas. 

In the remainder of this paper, we describe data we used, then present the archival UV and X-ray data analysis. This is followed by the study of DF44, and we close by discussing our results and summarizing our conclusions. 


\begin{table*}
\centering
\caption{Galaxies and archival UV and X-ray data used in this paper. \label{tab:datasources}}
\begin{tabular*}{\textwidth}{@{\extracolsep{\fill}} lcccclrrlcr}
\hline
{Name} & R.A. & Dec. & $r_{\text{eff}}$ & $R_{\text{Coma}}$ & GALEX tile & $t_{\text{FUV}}$ & $t_{\text{NUV}}$ & \textit{Chandra} & Chip & $t_{\text{X-ray}}$ \\
       &      &      &                  &                   &            &                  &     & obsID  & & \\
 & [J2000] & [J2000] & [kpc] & [Mpc] & & [ks] & [ks] &  &  [ks] \\
\hline
DF1  & 12:59:14.1 & 29:07:16	& 3.1	& 2.0 & GI1\_039006\_Coma\_MOS06    & 1.7   & 3.5  & \cds & \cds & \cds\\
DF2	 & 12:59:09.5 & 29:00:25	& 2.1	& 1.8 & GI1\_039006\_Coma\_MOS06    & 1.7   & 3.5  & \cds & \cds & \cds\\
DF3	 & 13:02:16.5 & 28:57:17	& 2.9	& 1.9 & GI1\_039006\_Coma\_MOS06    & 1.7   & 2.6  & \cds & \cds & \cds\\
DF4	 & 13:02:33.4 & 28:34:51	& 3.9	& 1.5 & GI1\_039008\_Coma\_MOS08    & 1.7   & 2.6  & \cds & \cds & \cds\\
DF5	 & 12:55:10.5 & 28:33:32	& 1.8	& 2.0 & GI1\_039003\_Coma\_MOS03    & 1.7   & 2.5  & \cds & \cds & \cds\\
DF6	 & 12:56:29.7 & 28:26:40	& 4.4	& 1.5 & GI1\_039003\_Coma\_MOS03    & 1.7   & 2.5  & \cds & \cds & \cds\\
DF7	 & 12:57:01.7 & 28:23:25	& 4.3	& 1.3 & GI1\_039003\_Coma\_MOS03    & 1.7   & 2.5  & \cds & \cds & \cds\\
DF8	 & 13:01:30.4 & 28:22:28	& 4.4	& 0.9 & GI5\_025001\_COMA           & 18.6  & 25.7 & \cds & \cds & \cds\\
DF9	 & 12:56:22.8 & 28:19:53	& 2.8	& 1.5 & GI1\_039003\_Coma\_MOS03    & 1.7   & 2.5  & \cds & \cds & \cds\\
DF10 & 12:59:16.3 & 28:17:51	& 2.4	& 0.6 & GI5\_025001\_COMA           & 18.6  & 25.7 & \cds & \cds & \cds\\
DF11 & 13:02:25.5 & 28:13:58    & 2.1   & 1.1 & GI5\_025001\_COMA           & 18.6  & 25.7 & \cds & \cds & \cds\\
DF12 & 13:00:09.1 & 28:08:27	& 2.6	& 0.3 & GI5\_025001\_COMA           & 18.6  & 25.7 & 18235 & ACIS-S & 30 \\
DF13 & 13:01:56.2 & 28:07:23	& 2.2	& 0.9 & GI5\_025001\_COMA           & 18.6  & 25.7 & \cds & \cds & \cds\\
DF14 & 12:58:07.8 & 27:54:46	& 3.8	& 0.7 & GI5\_025001\_COMA           & 18.6  & 25.7 & 18236, 19909, & ACIS-I & 30 \\ 
     &            &             &       &     &                             &       &      & 19910 & & \\
DF15 & 12:58:16.3 & 27:53:29	& 4.0   & 0.6 & GI5\_025001\_COMA           & 18.6  & 25.7 & 18236, 19909, & ACIS-I & 30 \\
     &            &             &       &     &                             &       &      & 19910 & & \\
DF16 & 12:56:52.4 & 27:52:29	& 1.5	& 1.1 & COMA\_SPEC\_A               & 1.3   & 2.8  & \cds & \cds & \cds\\
DF17 & 13:01:58.3 & 27:50:11	& 4.4	& 0.9 & GI5\_025001\_COMA           & 18.6  & 25.7 & 10921 & ACIS-S & 5 \\
DF18 & 12:59:09.3 & 27:49:48	& 2.8	& 0.4 & GI5\_025001\_COMA           & 18.6  & 25.7 & 13994, 14411 & ACIS-I & 110 \\
DF19 & 13:04:05.1 & 27:48:05	& 4.4	& 1.7 & GI1\_039009\_Coma\_MOS09    & 1.7   & 1.7  & \cds & \cds & \cds\\
DF20 & 13:00:18.9 & 27:48:06	& 2.3	& 0.4 & GI5\_025001\_COMA           & 18.6  & 25.7 & 2941, 13993, & ACIS-I & 303 \\ 
     &            &             &       &     &                             &       &      & 14410, 13995, & & \\ 
     &            &             &       &     &                             &       &      & 14406, 14415 & & \\
DF21 & 13:02:04.1 & 27:47:55	& 1.5	& 0.9 & GI5\_025001\_COMA           & 18.6  & 25.7 & 10921 & ACIS-S & 5 \\
DF22 & 13:02:57.8 & 27:47:25	& 2.1	& 1.3 & GI1\_039009\_Coma\_MOS09    & 1.7   & 1.7  & \cds & \cds & \cds\\
DF23 & 12:59:23.8 & 27:47:27	& 2.3	& 0.4 & GI5\_025001\_COMA           & 18.6  & 25.7 & 13993, 14410, & ACIS-I & 233 \\
     &            &             &       &     &                             &       &      & 13994, 14411 &  & \\
DF24 & 12:56:28.9 & 27:46:19	& 1.8	& 1.3 & COMA\_SPEC\_A               & 1.3   & 2.8  & \cds & \cds & \cds\\
DF25 & 12:59:48.7 & 27:46:39	& 4.4	& 0.3 & GI5\_025001\_COMA           & 18.6  & 25.7 & 13993, 14410, & ACIS-I & 479 \\ 
     &            &             &       &     &                             &       &      & 13994, 14411, & & \\
     &            &             &       &     &                             &       &      & 13995, 14406, & & \\
     &            &             &       &     &                             &       &      & 14415, 13996 & & \\
DF26 & 13:00:20.6 & 27:47:13	& 3.3	& 0.4 & GI5\_025001\_COMA           & 18.6  & 25.7 & \cds & \cds & \cds\\
DF27 & 12:58:57.3 & 27:44:39    & \cds  & 0.5 & GI5\_025001\_COMA           & 18.6  & 25.7 & 18237, 19911, & ACIS-I & 30 \\
     &            &             &       &     &                             &       &      & 19912 & & \\
DF28 & 12:59:30.4 & 27:44:50	& 2.7	& 0.4 & GI5\_025001\_COMA           & 18.6  & 25.7 & 18237, 19911,  & ACIS-S & 30 \\
     &            &             &       &     &                             &       &      & 19912 & & \\
DF29 & 12:58:05.0 & 27:43:59	& 3.1	& 0.8 & GI5\_025001\_COMA           & 18.6  & 25.7 & \cds & \cds & \cds\\
DF30 & 12:53:15.1 & 27:41:15    & 3.2   & 2.6 & $\cdots$                    & \cds  & \cds & \cds & \cds & \cds\\
DF31 & 12:55:06.2 & 27:37:27	& 2.5	& 1.9 & COMA\_SPEC\_A               & 1.3   & 2.8  & \cds & \cds & \cds\\
DF32 & 12:56:28.4 & 27:37:06	& 2.8	& 1.4 & GI2\_046001\_COMA3          & 29.9  & 31.1 & \cds & \cds & \cds\\
DF33 & 12:55:30.1 & 27:34:50	& 1.9	& 1.8 & GI2\_046001\_COMA3          & 29.9  & 31.1 & \cds & \cds & \cds\\
DF34 & 12:56:12.9 & 27:32:52	& 3.4	& 1.6 & GI2\_046001\_COMA3          & 29.9  & 31.1 & \cds & \cds & \cds\\
DF35 & 13:00:35.7 & 27:29:51    & 2.7   & 0.9 & GI5\_025001\_COMA           & 18.6  & 25.7 & \cds & \cds & \cds\\
DF36 & 12:55:55.4 & 27:27:36	& 2.6	& 1.8 & GI2\_046001\_COMA3          & 29.9  & 31.1 & \cds & \cds & \cds\\
DF37 & 12:59:23.6 & 27:21:22    & 1.5   & 1.1 & GI2\_046001\_COMA3          & 29.9  & 31.1 & \cds & \cds & \cds\\
DF38 & 13:02:00.1 & 27:19:51	& 1.8	& 1.4 & GI1\_039009\_Coma\_MOS09    & 1.7   & 1.7  & \cds & \cds & \cds\\
DF39 & 12:58:10.4 & 27:19:11	& 4.0   & 1.3 & GI2\_046001\_COMA3          & 29.9  & 31.1 & 4724 & ACIS-I & 60 \\
DF40 & 12:58:01.1 & 27:11:26	& 2.9	& 1.5 & GI2\_046001\_COMA3          & 29.9  & 31.1 & \cds & \cds & \cds\\
DF41 & 12:57:19.0 & 27:05:56	& 3.4	& 1.8 & GI2\_046001\_COMA3          & 29.9  & 31.1 & \cds & \cds & \cds\\
DF42 & 13:01:19.1 & 27:03:15	& 2.9	& 1.7 & GI5\_063019\_A1656\_FIELD2  & 1.5   & 1.6  & \cds & \cds & \cds\\
DF43 & 12:54:51.4 & 26:59:46	& 1.5	& 2.6 & GI2\_046001\_COMA3          & 29.9  & 31.1 & \cds & \cds & \cds\\
DF44 & 13:00:58.0 & 26:58:35	& 4.6	& 1.8 & GI5\_063019\_A1656\_FIELD2  & 1.5   & 4.2  & \cds & \cds & \cds\\
DF45 & 12:53:53.7 & 26:56:48	& 1.9	& 2.9 & GA\_DDO154                  & 4.2   & 4.2  & \cds & \cds & \cds\\
DF46 & 13:00:47.3 & 26:46:59	& 3.4	& 2.1 & GI5\_063019\_A1656\_FIELD2  & 1.5   & 2.6  & \cds & \cds & \cds\\
DF47 & 12:55:48.1 & 26:33:53	& 4.2	& 2.9 & GI1\_039004\_Coma\_MOS04    & 1.7   & 2.5  & \cds & \cds & \cds\\
\hline
\multicolumn{11}{p{0.9\textwidth}}{Columns. (1) Dragonfly ID from \cite{vdokkum2015}, (2) Right Ascension, (3) Declination, (4) Effective radius, (5) projected distance from Coma center, (6) GALEX observation used, (7-8) NUV and FUV exposure times, (9) \textit{Chandra} observations used, (10) \textit{Chandra} detector used, (11) \textit{Chandra} exposure time.}
\end{tabular*}
\end{table*}

\section{Observations and Data Processing}

The Coma DF galaxies, archival UV observations, and archival X-ray observations used in this paper are given in Table~\ref{tab:datasources}. The measurements and upper limits from these data are given in Table~\ref{tab:udgoptuvxray}. 

\subsection{GALEX}

Most of the DF survey footprint is covered by deep GALEX NUV and FUV exposures, and they tend to be the brighter and less confused sources within the Coma cluster, which motivates our use of the DF sample. Nevertheless, none of the DF galaxies are in the GALEX source catalog, so we performed aperture photometry with Photutils v0.5 \citep{bradley2017} on calibrated images retrieved from MAST\footnote{The Mikulski Archive for Space Telescopes;  https://archive.stsci.edu}. For each individual galaxy, we centered an aperture {at the DF catalog coordinates with a radius equal to the optical half-light radius and apply the aperture correction.} We subtracted the local background after masking point sources (detected at $2.5\sigma$) and sigma-clipping the background (removing deviations at $\sigma > 2.5$). We visually inspected each source to verify this procedure. {Out of the 47 DF targets, 7 were rejected in the FUV channel and 8 were rejected in the NUV channel due to additional sources in the aperture.} We used the same images for stacking by clipping the image to a region several times the effective radius in size and subtracted the estimated background.

At a 3$\sigma$ threshold, we detected {3/40 DF UDGs in the FUV channel and 6/39 in the NUV} (Table~\ref{tab:udgoptuvxray}). This is {inconsistent with} \citet{singh2019}, who did not detect any DF UDGs in Coma and only 2/110 of the SMUDGes UDGs within the Coma virial radius were detected. They also used Photutils on the same data, with the main difference being that they measured the sky background within the same aperture as the UDG from the GALEX-provided sky background image. They verified that this heavily smoothed background map was not contaminated by the UDG itself. However, their method of identifying detections is different: whereas we use the 3$\sigma$ threshold based on the sigma-clipped background, they classify sources with $>2\sigma$ as candidate detections, then visually inspect the data to determine whether the candidate is consistent with the UDG position or affected by contamination. 

Most of our UV-detected sources are marginal detections ($3< \sigma < 5$), so on a case-by-case basis it is often unclear whether the source is mis-centered, contaminated, etc. However, the stacked image of all the sources  (Figure~\ref{fig:fullstackims}) shows a very strong signal from the NUV channel, which gives us confidence that at least some Coma DF UDGs do have UV emission; DF44 is separately described below. 

\subsection{SDSS}

The UV-optical color is of interest to determine the stellar populations of UDGs, so we used \textit{ugriz} data from the Sloan Digital Sky Survey (SDSS) Data Release 12 \citep{alam2015}. Following the same procedure as with GALEX, we find the following detection fractions: {\textit{u} = 7/40, \textit{g} = 25/41, \textit{r} = 32/41, \textit{i} = 24/40, \textit{z} = 8/41.} For reference, \cite{york2000} provides 5$\sigma$ point-source detection limits for 1" seeing at an airmass of 1.4 of 22.3, 24.4, 24.1, 22.3, and 20.8 mags in the \textit{u, g, r, i} and \textit{z} filters, respectively, but with somewhat lower sensitivities for extended sources. The magnitudes and limits are reported in Table~\ref{tab:udgoptuvxray}. 

\subsection{\textit{Chandra}}

Most of the DF galaxies are in regions where intracluster X-ray emission from Coma is a major or dominant source of ``background'' for potential UDG emission from X-ray binaries or AGNs. This background is irreducible, so high angular resolution provides the best opportunity to detect UDGs in the X-rays and we used archival \textit{Chandra} data to search for UDG emission. 12 UDGs lie in \textit{Chandra} exposures. We obtained the archival data and re-processed it with the \textit{Chandra} Interactive Analysis of Observations (CIAO) v.4.8.2 software, following the standard pipeline. 

We filtered event files to 2-7~keV to optimize our sensitivity to nonthermal emission and merged overlapping files when the detector was the same and the observational parameters were similar, then performed aperture photometry using CIAO tools and local (annular) background. Unlike with the optical data, we used a single aperture size with $r=15^{\prime\prime}$ because of the stochastic distribution of bright X-ray binaries. No sources were detected, with 0.3-10~keV 3$\sigma$ upper limits of 1$-$7~$\times~10^{40}$~erg~s$^{-1}$ (Table~\ref{tab:udgoptuvxray}), assuming a power law with $\Gamma=1.8$ and a Galactic absorbing column from the Leiden-Argentine-Bonn \citep{kalberla05} survey\footnote{\url{https://heasarc.gsfc.nasa.gov/cgi-bin/Tools/w3nh/w3nh.pl}} of $7.98\times10^{19}$~cm$^{-2}$.

\begin{table*}
\centering
\caption{Dragonfly Coma X-ray Measurements. \label{tab:xray}}
\begin{tabular*}{\textwidth}{@{\extracolsep{\fill}} lccccccccc}
\hline
{Name} & $R_{\text{Coma}}$ & {Instrument} & {$t_{\text{exp}}$} & {Counts} & {$N_{\text{pix}}$} & $\bar{B}$ & Net Rate & 3$\sigma$ Sens. & $L_X$\\
       & [Mpc]             &              & [ks] &  &  & [Counts/pix] & [$10^{-4}$ Counts/s] & [$10^{-4}$ Counts/s] & [$10^{40}$ erg~s$^{-1}$] \\
\hline
DF12 & 0.3 & ACIS-S & 29.7  & 50  & 2922 & 0.014 & 2.8  & 6.5  & $<$3.2 \\
DF14 & 0.7 & ACIS-I & 27.4  & 31  & 2922 & 0.010 & 0.4  & 6.0  & $<$3.4 \\
DF15 & 0.6 & ACIS-I & 27.4  & 25  & 2922 & 0.010 & -1.7 & 6.0  & $<$3.4 \\
DF17 & 0.9 & ACIS-S & 5.0   & 7   & 2922 & 0.002 & 4.1  & 13.4 & $<$6.6 \\
DF18 & 0.4 & ACIS-I & 187.1 & 493 & 2922 & 0.161 & 1.2  & 3.5  & $<$2.0 \\
DF20 & 0.4 & ACIS-I & 158.8 & 310 & 2922 & 0.094 & 2.2  & 3.1  & $<$1.5 \\
DF21 & 0.9 & ACIS-S & 5.0   & 6   & 2922 & 0.002 & 2.1  & 13.4 & $<$7.6 \\
DF23 & 0.4 & ACIS-I & 43.5  & 155 & 2922 & 0.053 & 0.03 & 8.6  & $<$4.2 \\
DF25 & 0.3 & ACIS-I & 235.2 & 583 & 2922 & 0.203 & -0.4 & 3.1  & $<$1.8 \\
DF27 & 0.5 & ACIS-S & 265.9 & 531 & 2922 & 0.166 & 1.7  & 2.5  & $<$1.4 \\
DF28 & 0.4 & ACIS-S & 29.7  & 47  & 2922 & 0.016 & -0.2 & 7.0  & $<$4.0 \\
DF39 & 1.3 & ACIS-S & 59.7  & 28  & 2922 & 0.008 & 0.8  & 2.4  & $<$1.2 \\
DF44 & 1.8 & EPIC   & 166   & 143 & 143  & 1.23  & 0.02 & 0.5  & $<$0.01 \\
\hline
\multicolumn{10}{p{0.9\textwidth}}{Columns. (1) Dragonfly ID from \cite{vdokkum2015}, (2) projected distance from Coma cluster center, (3) Instrument (only DF44 is \xmm), (4) Vignetting-corrected exposure time at DF position, (5) Counts in aperture (2-7~keV for \textit{Chandra} observations, 0.4-2~keV for DF44), (6) Pixels in aperture, (7) Average background measured from annulus, (8) Net count rate, (9) 3$\sigma$ sensitivity count rate, (10) Limiting 0.3-10~keV $L_X$ (no galaxy was detected), based on a power law and the measurement bandpass.}\\
\multicolumn{10}{p{0.9\textwidth}}{Notes. Source fluxes were measured in circular apertures and backgrounds in surrounding annuli (see text).}
\end{tabular*}
\end{table*}

\subsection{DF44 \textit{Swift} Data}

We obtained about 90~ks of new \textit{Swift} UV and Optical Telescope (UVOT) and X-ray Telescope (XRT) data for DF44. The UVOT data are in the UVW2 (1928\AA) and UVW1 (2600\AA) bands (40~ks and 20~ks of good time, respectively), while the XRT data covers 0.3-10~keV. The UVOT data are accumulated as snapshots, so we combined them into a merged exposure after reduction using the standard pipeline, except to remove a diffuse artifact (described below). We also specified parameters in the pipeline to standardize the pixel scale, file format, and astrometric solution (known to $<0.1^{\prime\prime}$). Following the pipeline reduction, but before merging, we applied corrections based on the large-scale sensitivity maps (produced through the pipeline but not applied). 

Scattered light from the filter window creates persistent patterns on UVOT images. Since the UVOT operates by taking multiple snapshots for each observation and the pattern is static in detector coordinates \citep{breeveld11}, stacking images to improve sensitivity also stacks these artifacts with various offsets, which reduces the final sensitivity of the combined image. To mitigate this, we used templates for the scattered light in the UVW1 and UVW2 filters created by \citet{hk14} to subtract the artifact from each exposure. We masked the sources in each exposure down to a threshold of $3\sigma$ above the background, then used least-squares fitting to determine the appropriate scale factor for the template, which was then subtracted from the image. This technique can remove up to 95\% of the artifact, but can fail in crowded fields or with very short exposures, so the scale factors were verified by visual inspection (e.g., over-subtraction leads to a clear negative imprint of the features on the image). In cases where no solution was found, the image was not added to the stack {(about 33\% of the total UVW1 exposure time and 32\% of the total UVW2 exposure time)}. For comparison, we also stacked the images with no correction.

We then used aperture photometry to measure the UVW1 and UVW2 magnitudes for DF44 using the UVOT software. The UVW1 flux densities measured when correcting for the diffuse artifact and combining images without this correction are both $F_{\lambda} = 9\pm2\times 10^{-18}$~erg~s$^{-1}$~cm$^{-2}$~\AA$^{-1}$. The UVW2 corrected and uncorrected fluxes are $F_{\lambda} = 6\pm3\times 10^{-18}$~erg~s$^{-1}$~cm$^{-2}$~\AA$^{-1}$ and $3\pm2\times 10^{-18}$~erg~s$^{-1}$~cm$^{-2}$~\AA$^{-1}$. These values are consistent with each other. 

Finally, the fluxes must be corrected for a red transmission leak. In the UVW1 filter there is a small but non-negligible response between 3000-5500\AA, so that a significant fraction of the measured UVW1 flux can be optical light in red sources. DF44 appears to have optical flux densities several times higher than the UV flux densities, so this cannot be ignored. 

We used the SDSS $u$ (2980-4130\AA) and $g$ (3630-5680\AA) filters along with the UVW1 transmission curve to estimate the optical contamination, based on the following expression (using fluxes rather than flux densities):
\begin{equation*}
    F_{\textrm{leak,SDSS}} = \bigg(\frac{\int_{\lambda_{\textrm{min,SDSS}}}^{\lambda_{\textrm{max,SDSS}}}T_{UVW1}}{\int_{\lambda_{\textrm{min},UVW1}}^{\lambda_{\textrm{max},UVW1}}T_{UVW1}}\bigg) \times F_{\textrm{SDSS}}[\textrm{erg}~\textrm{s}^{-1} \textrm{cm}^{-2}].
\end{equation*}
where $F_{\textrm{leak,SDSS}}$ is the flux measured in the UVW1 band from only that flux seen in the \textit{u} or \textit{g} SDSS filter, and can be subtracted from the measured UVW1 flux to obtain the true UV flux, $F_{\textrm{corr},UVW1} = F_{UVW1} - F_{\textrm{leak},u} - F_{\textrm{leak},g}$. We then measured the effective wavelength of the ``corrected'' filter as 2505\AA. We estimate that the optical contribution to the DF44 UVW1 flux is 14\%, so its true UV flux near 2500\AA\ is $7.2\times10^{-18}$ erg~s$^{-1}$cm$^{-2}$\AA$^{-1}$; 14\% is comparable to the flux uncertainty. After subtraction, the UVW1 flux density is closer to that in the NUV, which is expected based on the overlap of the transmission curves. The effective wavelengths, flux densities, uncertainties, exposure times, and conversion factors for all relevant filters used for DF44 (including GALEX and SDSS) are reported in Table \ref{tab:df44phot}. 

The XRT data were obtained in photon-counting mode and were reduced using the HEASoft XRT analysis tools\footnote{\url{https://swift.gsfc.nasa.gov/analysis/}}. The total good time was 88~ks, and we combined the exposures in \textit{Xselect}\footnote{\url{https://heasarc.nasa.gov/ftools/xselect/}} and created a corrected exposure map. For analysis we used the stacked images in both the 0.3-2~keV and 0.3-10~keV bandpasses. No source is detected at the position of DF44, with a 0.3-10~keV upper limit of $L_X < 4.1\times 10^{39}$~erg~s$^{-1}$, assuming a power-law with $\Gamma=1.8$ and Galactic absorption.  

\begin{table*}
\centering
\caption{Dragonfly Coma UDG UV, Optical, and X-ray measurements. \label{tab:udgoptuvxray}}
\begin{tabular*}{\textwidth}{@{\extracolsep{\fill}} lccccccccc}
\hline
{Name} & $R_{\text{Coma}}$ & $m_{\text{FUV}}$ & $m_{\text{NUV}}$ & $m_u$ & $m_g$    & $m_r$             & $m_i$             & $m_z$             & $L_X$ \\
     & [Mpc]& [mag]         & [mag]          & [mag]            & [mag]             & [mag]             & [mag]             & [mag]             & [erg s$^{-1}$] \\
\hline
DF1  & 2.0 & $>$24.47       & $>$24.29       & $>$20.90         & 20.22$\pm$0.15    & 19.94$\pm$0.20    & 19.76$\pm$0.26    & $>$19.41          & \cds \\
DF2	 & 1.8 & $>$25.77       & $>$24.26       & $>$21.06         & $>$22.35          & $>$21.76          & $>$21.13          & $>$19.87          & \cds \\
DF3	 & 1.9 & $>$25.33       & $>$24.11       & $>$20.47         & 20.28$\pm$0.17    & 19.59$\pm$0.14    & $>$20.71          & $>$19.33          & \cds \\
DF4	 & 1.5 & $>$24.85       & $>$23.50       & $>$20.68         & $>$21.27          & 19.78$\pm$0.21    & $>$20.53          & 18.21$\pm$0.31    & \cds \\
DF5	 & 2.0 & $>$26.08       & $>$24.28       & $>$21.29         & $>$22.08          & 20.55$\pm$0.21    & 20.24$\pm$0.32    & $>$19.85          & \cds \\
DF6	 & 1.5 & $>$24.12       & $>$23.39       & \cds             & $>$21.44          & 19.81$\pm$0.24    & $>$20.32          & $>$18.86          & \cds \\
DF7	 & 1.3 & $>$24.58       & 22.47$\pm$0.30 & 19.06$\pm$0.21   & 19.16$\pm$0.09    & 18.42$\pm$0.07    & 18.31$\pm$0.09    & $>$18.87          & \cds \\
DF8	 & 0.9 & $>$24.69       & $>$24.34       & $>$20.28         & 20.20$\pm$0.22    & 19.35$\pm$0.16    & $>$20.34          & $>$18.88          & \cds \\
DF9	 & 1.5 & $>$24.97       & $>$23.98       & 19.81$\pm$0.28   & 20.33$\pm$0.17    & $>$21.43          & 19.83$\pm$0.28    & $>$19.36          & \cds \\
DF10 & 0.6 & $>$25.44       & 24.48$\pm$0.31 & $>$20.97         & 20.93$\pm$0.25    & 19.49$\pm$0.11    & 19.37$\pm$0.16    & 18.45$\pm$0.25    & \cds \\
DF12 & 0.3 & $>$25.41       & $>$25.97       & $>$21.05         & 20.96$\pm$0.22    & 20.31$\pm$0.22    & 20.11$\pm$0.29    & $>$19.56          & $<3.2\times10^{40}$ \\
DF13 & 0.9 & $>$25.51       & $>$25.39       & $>$21.23         & $>$21.1           & $>$21.75          & $>$21.23          & $>$19.81          & \cds \\
DF14 & 0.7 & $>$24.72       & $>$24.32       & $>$20.52         & 20.23$\pm$0.20    & $>$21.11          & 19.61$\pm$0.31    & 18.18$\pm$0.30    & $<3.4\times10^{40}$\\ 
DF15 & 0.6 & $>$24.81       & $>$24.61       & $>$20.65         & 19.74$\pm$0.14    & 19.48$\pm$0.16    & 19.40$\pm$0.27    & 17.61$\pm$0.18    & $<3.4\times10^{40}$ \\
DF16 & 1.1 & $>$25.33       & $>$25.27       & $>$21.62         & $>$22.69          & 21.14$\pm$0.28    & \cds              & \cds              & \cds \\
DF17 & 0.9 & $>$24.79       & $>$24.32       & $>$20.52         & 19.16$\pm$0.09    & 19.66$\pm$0.21    & 18.66$\pm$0.16    & 18.12$\pm$0.33    & $<6.6\times10^{40}$ \\
DF18 & 0.4 & 24.01$\pm$0.28 & $>$24.99       & $>$20.97         & $>$21.92          & 20.38$\pm$0.27    & $>$20.70          & $>$19.37          & $<2.0\times10^{40}$ \\
DF19 & 1.7 & $>$24.64       & $>$23.38       & $>$20.40         & $>$21.44          & $>$20.95          & $>$20.24          & $>$18.91         & \cds \\
DF20 & 0.4 & $>$25.23       & $>$24.89       & $>$21.22         & $>$22.13          & 20.80$\pm$0.32    & $>$20.96          & $>$19.66          & $<1.5\times10^{40}$ \\ 
DF21 & 0.9 & $>$26.11       & $>$26.09       & $>$21.60         & 20.50$\pm$0.13    & 20.13$\pm$0.12    & 19.90$\pm$0.17    & $>$20.01          & $<7.6\times10^{40}$ \\
DF22 & 1.3 & \cds           & $>$24.44       & $>$21.39         & $>$21.1           & 20.24$\pm$0.18    & 19.88$\pm$0.22    & $>$19.73          & \cds \\
DF23 & 0.4 & $>$25.37       & 24.40$\pm$0.32 & $>$21.25         & 20.49$\pm$0.16    & 20.54$\pm$0.25    & 19.93$\pm$0.26    & $>$19.69          & $<4.2\times10^{40}$ \\
DF24 & 1.3 & $>$24.73       & $>$24.37       & $>$21.53         & $>$22.48          & $>$21.94          & $>$21.30          & $>$20.09          & \cds \\
DF25 & 0.3 & $>$24.70       & 23.12$\pm$0.16 & $>$20.55         & 20.10$\pm$0.20    & 19.94$\pm$0.27    & 19.32$\pm$0.29    & $>$18.92          & $<1.8\times10^{40}$ \\ 
DF26 & 0.4 & $>$24.78       & $>$24.94       & $>$20.81         & 19.46$\pm$0.10    & 18.69$\pm$0.07    & 18.44$\pm$0.10    & 18.13$\pm$0.25    & \cds \\
DF27 & 0.5 & \cds           & \cds           & \cds             & \cds              & \cds              & \cds              & \cds              & $<1.4\times10^{40}$ \\
DF28 & 0.4 & $>$25.40       & 23.55$\pm$0.16 & $>$21.02            & 20.24$\pm$0.14    & 19.32$\pm$0.09    & 19.48$\pm$0.18    & $>$19.62          & $<4.0\times10^{40}$ \\
DF29 & 0.8 & 23.94$\pm$0.26 & $>$25.07       & 19.96$\pm$0.28   & 20.39$\pm$0.18    & 19.59$\pm$0.14    & 19.59$\pm$0.23    & $>$19.50           & \cds \\
DF31 & 1.9 & \cds           & $>$23.95       & $>$20.75         & 20.79$\pm$0.23    & 20.26$\pm$0.23    & 19.80$\pm$0.24    & $>$19.15          & \cds \\
DF32 & 1.4 & $>$25.49       & \cds           & $>$20.65         & 20.26$\pm$0.16    & 19.44$\pm$0.12    & 19.87$\pm$0.28    & $>$19.03          & \cds \\
DF33 & 1.8 & $>$26.15       & \cds           & $>$21.18         & $>$22.37          & $>$21.88          & $>$21.28          & $>$19.48          & \cds \\
DF34 & 1.6 & $>$25.25       & \cds           & $>$20.57         & $>$21.75          & 20.30$\pm$0.29    & $>$20.68          & $>$18.85          & \cds \\
DF36 & 1.8 & $>$25.54       & \cds           & $>$20.81         & 20.54$\pm$0.19    & 20.00$\pm$0.19    & $>$20.84          & $>$19.24          & \cds \\
DF38 & 1.4 & $>$25.97       & $>$24.27       & $>$21.23         & 20.75$\pm$0.17    & 20.03$\pm$0.14    & $>$21.24          & $>$19.57          & \cds \\
DF39 & 1.3 & 23.97$\pm$0.23 & 23.43$\pm$0.17 & 19.55$\pm$0.27   & 20.59$\pm$0.26    & 19.40$\pm$0.14    & 19.04$\pm$0.16    & $>$18.76          & $<1.2\times10^{40}$ \\
DF40 & 1.5 & \cds           & \cds           & \cds             & \cds              & \cds              & \cds              & \cds              & \cds \\
DF41 & 1.8 & $>$25.05       & $>$24.81       & $>$20.69         & 20.52$\pm$0.22    & 19.67$\pm$0.17    & 19.34$\pm$0.20    & 18.03$\pm$0.26    & \cds \\
DF42 & 1.7 & $>$24.55       & $>$23.86       & $>$20.84         & $>$21.90          & 19.91$\pm$0.18    & 19.56$\pm$0.21    & $>$19.24          & \cds \\
DF43 & 2.6 & \cds	        & $>$26.42       & $>$21.75         & $>$22.72          & $>$22.19          & $>$21.65          & $>$20.14          & \cds \\
DF44 & 1.8 & $>$24.54       & $>$23.78       & 19.31$\pm$0.26   & 19.21$\pm$0.10    & 18.75$\pm$0.10    & 18.51$\pm$0.14    & $>$18.76          & $<1.1\times 10^{38}$ \\
DF45 & 2.9 & $>$25.26       & $>$24.49       & $>$21.56         & 20.51$\pm$0.13    & 20.41$\pm$0.18    & 19.91$\pm$0.17    & $>$19.95          & \cds \\
DF46 & 2.1 & $>$24.84       & $>$23.50       & $>$20.91         & $>$21.84          & 19.59$\pm$0.15    & $>$20.67          & 18.24$\pm$0.26    & \cds \\
DF47 & 2.9 & $>$24.96       & $>$23.53       & 19.84$\pm$0.32   & 20.31$\pm$0.21    & $>$21.05          & 19.12$\pm$0.19    & $>$19.03          & \cds \\
\hline
\multicolumn{10}{p{0.9\textwidth}}{Columns. (1) Dragonfly ID from \cite{vdokkum2015}, (2) projected distance from Coma cluster center, (3-7) AB magnitudes or 3$\sigma$ upper limits, (8) 3$\sigma$ upper limits to the 0.3-10~keV X-ray luminosity.} \\
\multicolumn{10}{p{0.9\textwidth}}{Notes. DF11, DF27, DF30, DF35, and DF37 were excluded from the UV analysis due to contamination or artifacts. Individual filters in several other galaxies were excluded for the same reasons.}
\end{tabular*}
\end{table*}

\subsection{DF44 \textit{XMM-Newton} Data}

\begin{figure*}
 \includegraphics[width=\linewidth]{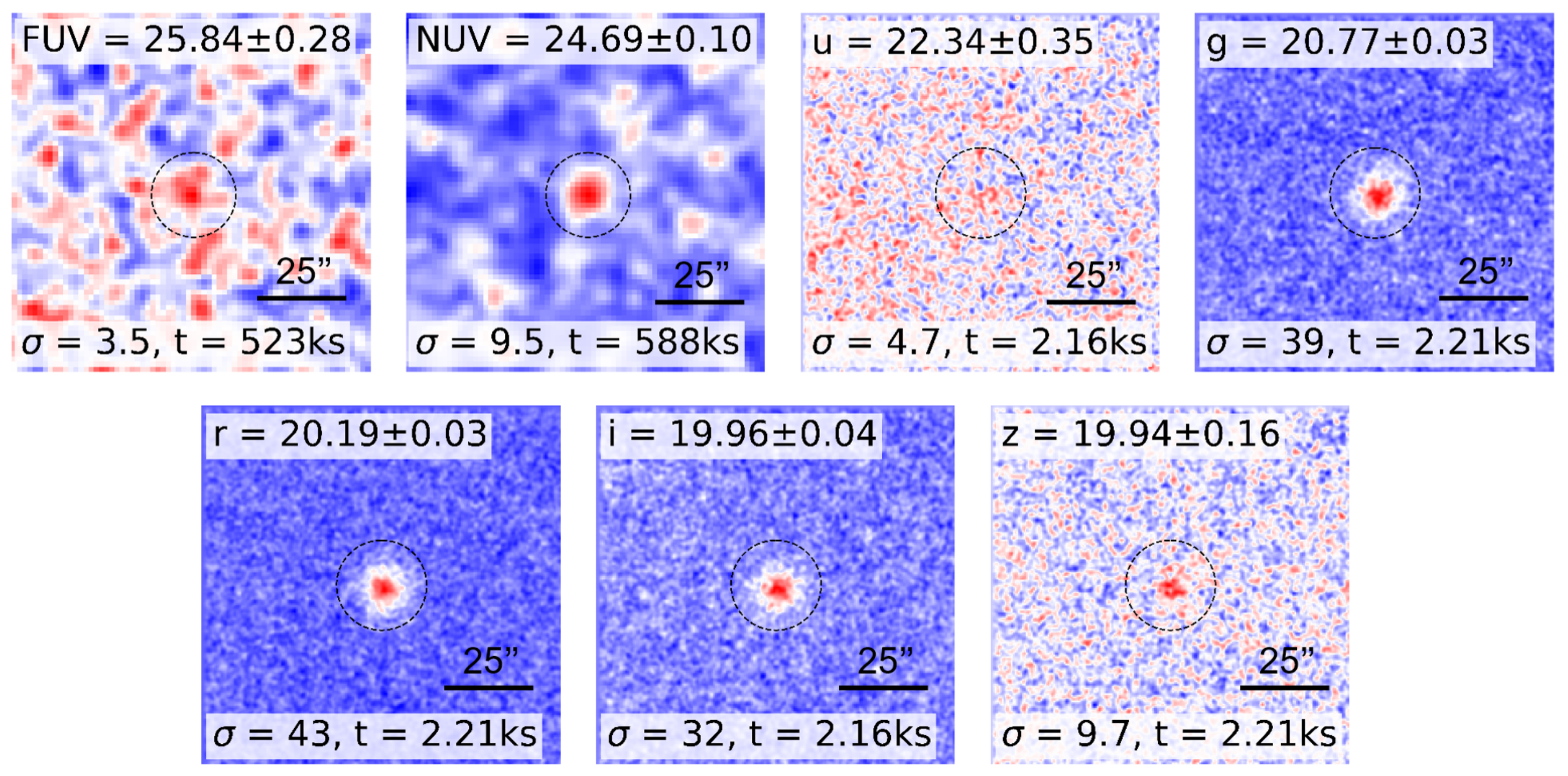}
 \caption{Full stacks of all DF data with filter, magnitude, source significance, and effective exposure time labeled. The images have been convolved with a Gaussian kernel of $\sigma=3$ pixels. The dashed black circle in each filter is 12" in radius.}
 \label{fig:fullstackims}
\end{figure*}

\begin{figure}
 \includegraphics[width=0.95\linewidth]{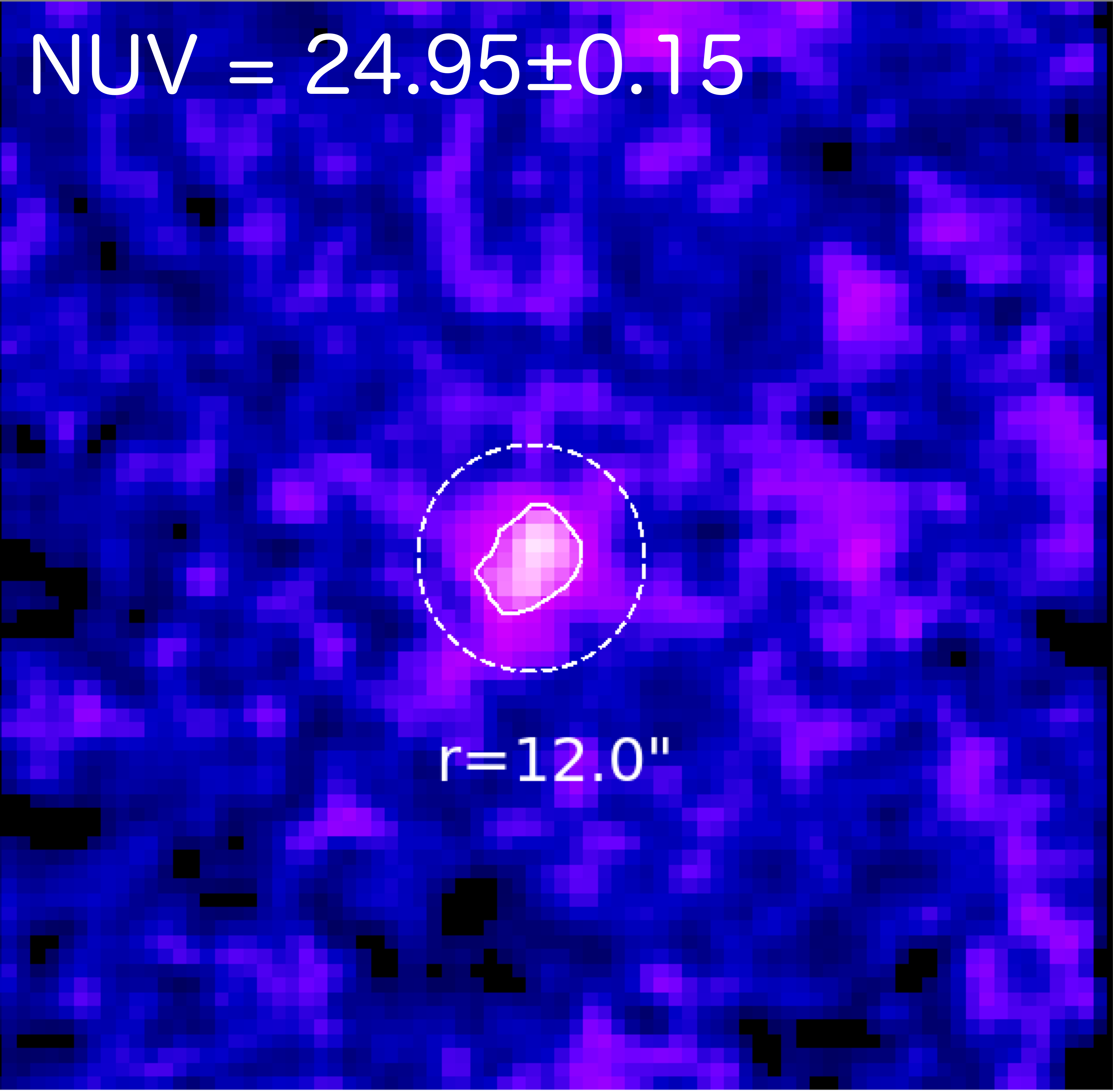}
 \caption{The stacked NUV image of all individually undetected DF galaxies shows a clear detection (3$\sigma$ contour above background shown in white). The apparent magnitude measured from within an aperture about twice the average effective radius is highly significant, motivating the use of shallower stacks in our analysis.}
 \label{fig:undetected_nuv}
\end{figure}

We obtained about 167~ks of new XMM observations, which were split between December 23rd, 2017 (OBS-ID:0800580101) and January 4th, 2018 (OBS-ID:0800580201). We used the EPIC MOS1, MOS2, and pn cameras. The XMM data were processed using standard methods with the Science Analysis System (SAS v16.0.0)\footnote{\url{https://www.cosmos.esa.int/web/xmm-newton/what-is-sas}}. We used the SAS tools {\tt emchain} and {\tt epchain} to register events for the MOS and pn cameras (we used events with pattern$\le$4), then measured the count rate over the full arrays in bins of 200~s to identify particle background flares, which were filtered out. After processing, the good time intervals (GTIs) for the three cameras in both observing periods are 66 ks and 46 ks in MOS1, 70 ks and 54 ks in MOS2, and 60 ks and 37 ks in pn, respectively. 

The quiescent particle background was subtracted based on the unexposed array corners with the Extended-SAS software. We then filtered the images to the 0.4-2~keV bandpass for source detection, which maximizes the sensitivity to the expected sources (hot gas or LMXBs). However, our results are consistent with measurements made from a 0.3-10~keV bandpass. No source is detected at the position of DF44, and the 0.3-10~keV, 3$\sigma$ upper limit of $L_X< 1.0\times 10^{38}$~erg~s$^{-1}$ is much lower than the XRT value ($L_X < 4.1\times 10^{39}$~erg~s$^{-1}$). 

DF44 is on the outskirts of the Coma cluster, just south of the mosaic produced by \citet{Briel2001}. In the XMM field we do not find strong cluster emission, and instead find that such emission can only be detected when integrating over large areas and that its surface brightness is below that of the other backgrounds and foregrounds. We find that the background around DF44 is essentially flat (when accounting for vignetting and the extended wings of other nearby sources) on a scale of several arcmin, so we adopt an annular region of width 10~arcsec for the background. By fitting a linear gradient to the 0.4--2.0~keV image, we estimate that the uncertainty in the upper limit quoted above is of order 1\% due to the intracluster medium, i.e., smaller than other uncertainties.

\section{Ultraviolet Fluxes and Colors}

A minority of galaxies were detected in either the NUV or FUV channels, and the detected magnitudes or upper limits are consistent with quiescent galaxies. To improve the sensitivity, we stacked the data in both GALEX and the \textit{ugriz} SDSS bands to measure the average SED (Figure~\ref{fig:fullstackims}; for presentation, each stack was convolved with a Gaussian kernel of $\sigma=3$ pixels). The SED is consistent with the DF UDGs being quiescent, red galaxies. However, in contrast to \citet{singh2019}, who argued that stacking was not likely to be useful for ``red'' UDGs, we detect some DF UDGs individually and more through stacks. Figure~\ref{fig:undetected_nuv} shows the stacked NUV galaxies that are \textit{not} individually detected. There is clearly a strong detection (the white contour is 3$\sigma$ above background).

\subsection{Correlation with Distance}

\begin{figure*}
 \includegraphics[width=\linewidth]{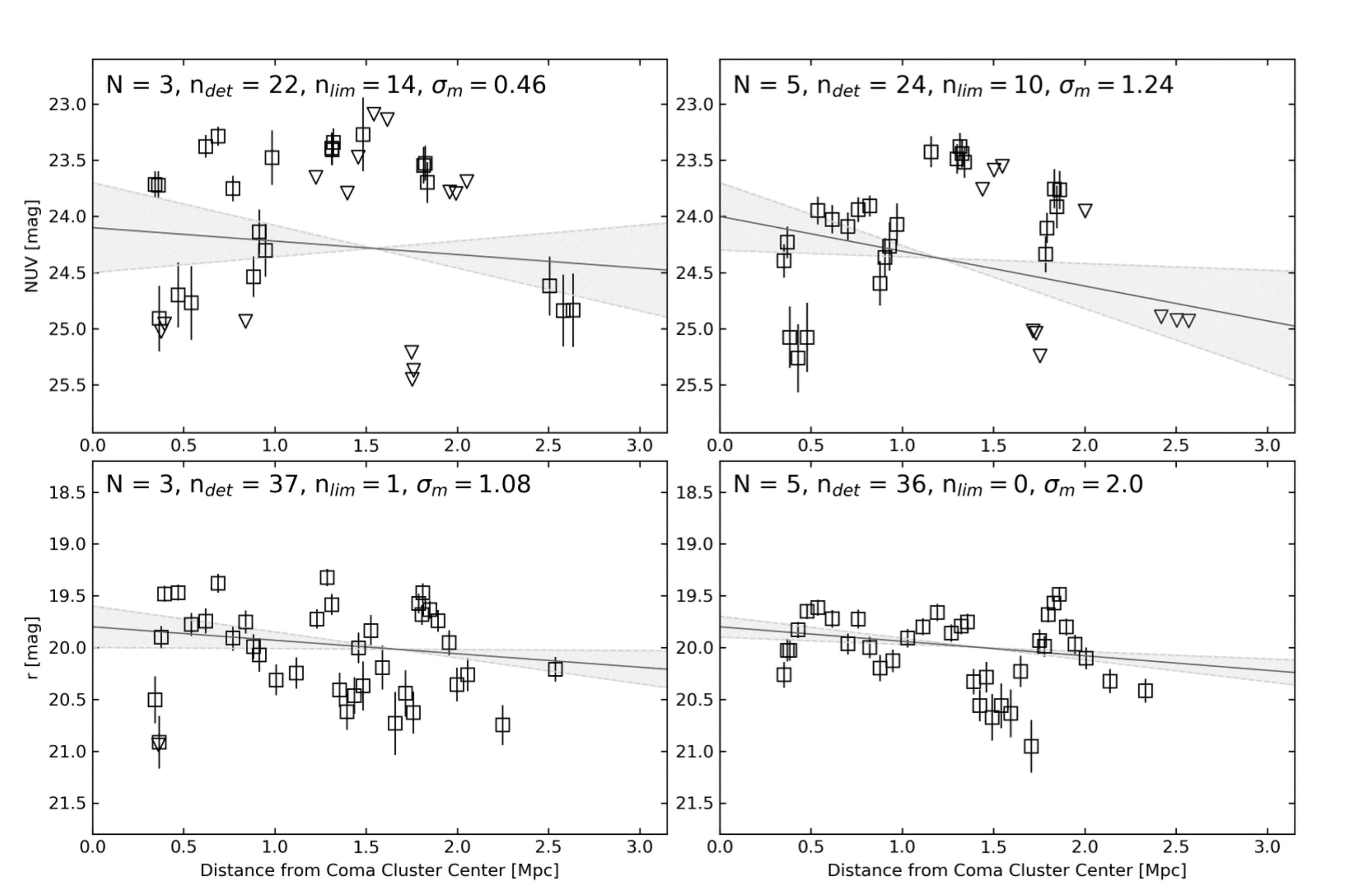}
 \caption{Scatter plots of apparent magnitudes from running stack samples for NUV and \textit{r} data as a function of the weighted average stack distance from the cluster center. All scatters were fit with \texttt{linmix}. N denotes the number of galaxies in each stack, $n_{\mathrm{det}}$ denotes the number of points with S/N $>$ 3 which are plotted by squares with error bars, $n_{\mathrm{lim}}$ denotes the number of points fit as $3\sigma$ upper limits which are plotted by inverted triangles, and $\sigma_{\mathrm{m}}$ denotes the slope significance. Note $\sigma_{\mathrm{m}}$ is different from the $\sigma$ indicating the Gaussian scatter in the regression. The lighter dashed grey lines indicate the slope uncertainty.}
 \label{fig:NUV_R_runningstack}
\end{figure*}

Motivated by the \cite{jiang2019} scenario in which ram-pressure stripping quenches star formation and the good signal in the full stacks, we also stacked sub-samples of the data to search for a correlation between UV luminosity and distance from the cluster core.  First, we created a ``running'' stack in which $N$ galaxies were stacked at a time, with the first point combining the $N$ galaxies closest to the core, and the next point excluding the closest galaxy and including the next farthest not already in the stack. Since the stacked images have different exposure times, the mean distance for each point is weighted by the exposures of each constituent. 

Running stacks with $N=3$ and 5 were created for the FUV, NUV, and SDSS $ugriz$ bands. Of these, the FUV, $u$, and $z$ bands lack the signal for useful fitting. We fitted the remainder using the {hierarchical Bayesian linear regression model} \citep[{\tt linmix} from][]{Kelly2007}, which has the form $m = \alpha + \beta R \pm \sigma$, where $m$ is the magnitude, $\alpha$ is the intercept, $\beta$ is the slope, $R$ is the projected radius from the Coma core in Mpc, and $\sigma$ is a Gaussian scatter term. {Upper limits are included through standard approaches to censored data.} The significance of each parameter is estimated from the posterior distribution produced by {\tt linmix}. {The $g$ and $i$ band fits are similar to $r$ with less significance, so only the fits to $r$ are presented.} The NUV and $r$ fits are summarized in Table~\ref{tab:stackfits}. The fits (especially to the $N=5$ stacks) hint at correlations with distance, but none are inconsistent with zero slope at the $3\sigma$ level. 

\begin{figure}
  \includegraphics[width=\linewidth]{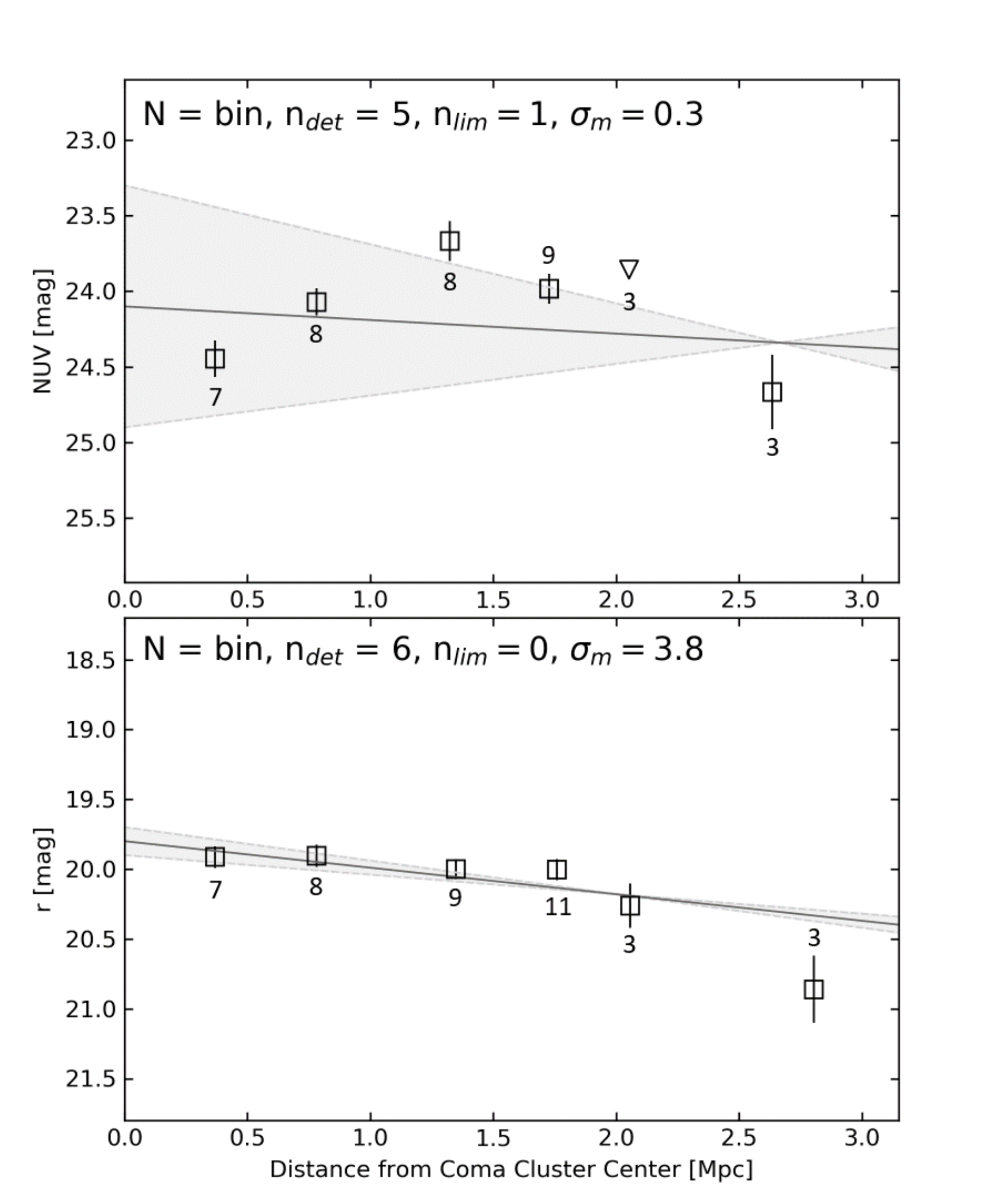}
 \caption{{ Scatter plots of apparent NUV (\textit{top}) and \textit{r} (\textit{bottom}) magnitudes from galaxy stacks of 0.5 Mpc bins with a similar layout as Figure \ref{fig:NUV_R_runningstack}. The numbers next to the data points denote the number of galaxies within each stack and triangles indicate upper limits. Both scatters were fit using \texttt{linmix}.}}
 \label{fig:NUV_R_binstack}
\end{figure}

In addition to the running stacks, we measured the fluxes from stacks based on distance (i.e., with a non-uniform number of galaxies per stack). We used bins 0.5~Mpc wide, as shown in Figure~\ref{fig:NUV_R_binstack}. In all but one NUV bin we have a clear detection, so we applied the same regression as to the running stacks. We find a significant correlation of $r$-band magnitude with distance but not NUV for this binning. The FUV is too faint to search for a correlation, and the signal-to-noise ratio in $g$, $i$, and $z$ stacks is considerably lower than for $r$, leading to best-fit slopes in agreement with $r$ but less significant.

The running stacks have an advantage over the distance-binned stack in that the distance-binned stack can have many more galaxies per data point at one distance than another. Since there is certainly scatter in the magnitudes at any radius, the formal uncertainties on the magnitudes in the distance-binned stack could be too low. However, the running stacks also have a disadvantage in that the central data points are each used $n=3$ or 5 times, whereas the outermost point only contributes to a single stack. The outermost galaxies provide a lever arm on any correlation and are important. We note that the reported slopes for each stack method are consistent with each other, supporting the prospect of a weak correlation of $r$ magnitude with distance and significant galaxy-to-galaxy scatter.

\subsection{UV-Optical Colors}

\begin{figure}
 \includegraphics[width=\linewidth]{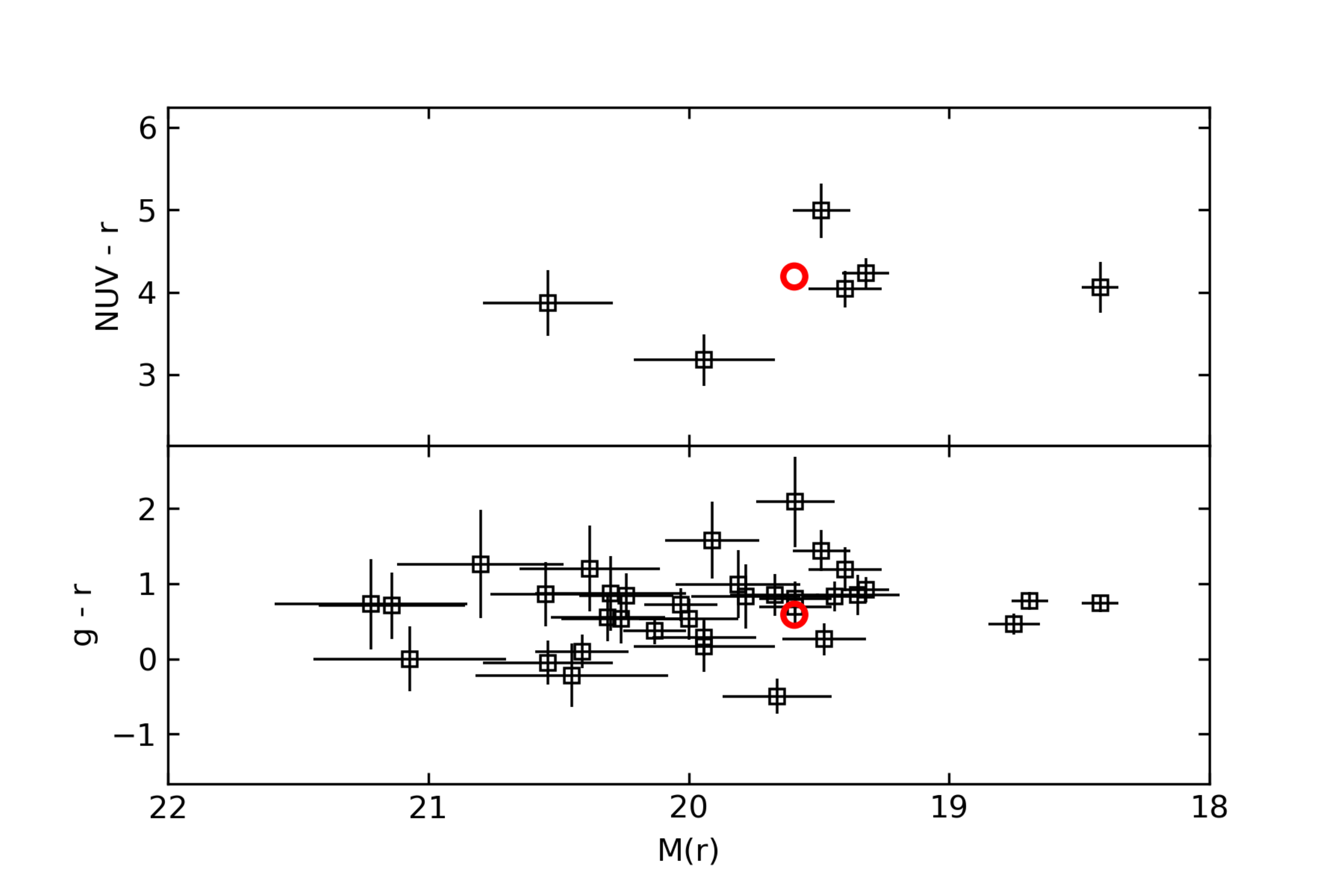}
 \caption{Color-magnitude comparisons of NUV$-$\textit{r} \textit{(top)} and \textit{g}$-$\textit{r} \textit{(bottom)} against \textit{r} magnitudes for the DF sample. The red circle denotes DF44's position in the plot. Only galaxies with S/N $>$ 3 are shown.}
 \label{fig:dfcmrcomparisons}
\end{figure}

The NUV detections suggest that the UDGs have recently formed stars. \citet{kaviraj2007} studied UV-optical colors of early-type galaxies and found that a color index of NUV$-r < 5.5$~mag indicates star formation within the past Gyr, with the younger stars representing $\sim$1\% of the galaxy's stellar mass (implying a mean SFR$< 10^{-3} M_{\odot}$~yr$^{-1}$). This is consistent with a similar analysis by \cite{Schawinski2007} that identified NUV$-r < 5.4$~mag as indicating recent star formation. Although this criterion was applied to early-type galaxies and not UDGs, one would expect UV-optical colors to reflect starlight and the UV-optical color-magnitude diagram for the UDGs (Figure~\ref{fig:dfcmrcomparisons}) is consistent with early-type galaxies \citep[see Figure~4 in][]{kaviraj2007}. Moreover, most of the Coma DF UDGs are part of the ``red'' family of UDGs \citep{zaritsky2019}. Notably, all of the UDGs fall below the NUV$-r < 5.5$~mag cutoff, so they have likely formed some stars in the past Gyr. This is also true for the average NUV$-r$ color for the full stack of about 4.7~mag. 

Meanwhile, the mean FUV magnitude $m_{\text{FUV}} = 25.84\pm0.28$ from a stack of all Coma UDGs indicates that current star formation is much lower. This corresponds to a FUV luminosity $\nu L_{\nu} \approx 4\times 10^{39}$~erg~s$^{-1}$. \citet{Calzetti2013} suggest a conversion factor of SFR~$=3.0\times 10^{-47} \times \nu L_{\nu} M_{\odot}$~yr$^{-1}$, yielding SFR~$\approx 1.2\times 10^{-7} M_{\odot}$~yr$^{-1}$. There is clearly variation well in excess of the 0.28~mag uncertainty, with some galaxies detected with $m_{\text{FUV}} < 24$ (Table~\ref{tab:udgoptuvxray}). Nevertheless, the typical UDG is not forming stars rapidly enough to explain its NUV$-r$ color in the steady state. 

\begin{table}
\centering
\caption{Stacked NUV and $r$ Fit Parameters.\label{tab:stackfits}}
\begin{tabular*}{0.5\textwidth}{@{\extracolsep{\fill}}ccccc}
  \hline
{Band} & {N} & {Slope} & {Intercept} & {Significance} \\
       &     &         & [mag]       & [$\sigma$] \\
\hline
Running &     &        &              & \\
\hline
$r$ & 3 & $0.13\pm0.12$  & $19.8\pm0.2$ & 1.1\\
$r$ & 5 & $0.14\pm0.07$  & $19.8\pm0.1$ & 2.0 \\
NUV & 3 & $0.12\pm0.26$ & $24.1\pm0.4$ & 0.5 \\
NUV & 5 & $0.31\pm0.25$ & $24.0\pm0.3$ & 1.2 \\
\hline
Distance &   &  &  & \\
\hline
$r$ & bin & $0.19\pm0.05$ & $19.8\pm0.1$ & 3.8 \\
NUV & bin & $0.09\pm0.30$ & $24.1\pm0.8$ & 0.3 \\ 
\hline
\multicolumn{5}{p{0.4\textwidth}}{Notes. The values given are from a linear fit to the data shown in Figure~\ref{fig:NUV_R_runningstack} and Figure~\ref{fig:NUV_R_binstack}.}
\end{tabular*}
\end{table}

\section{\textit{Chandra} Limits}

\begin{figure}
 \includegraphics[width=\linewidth]{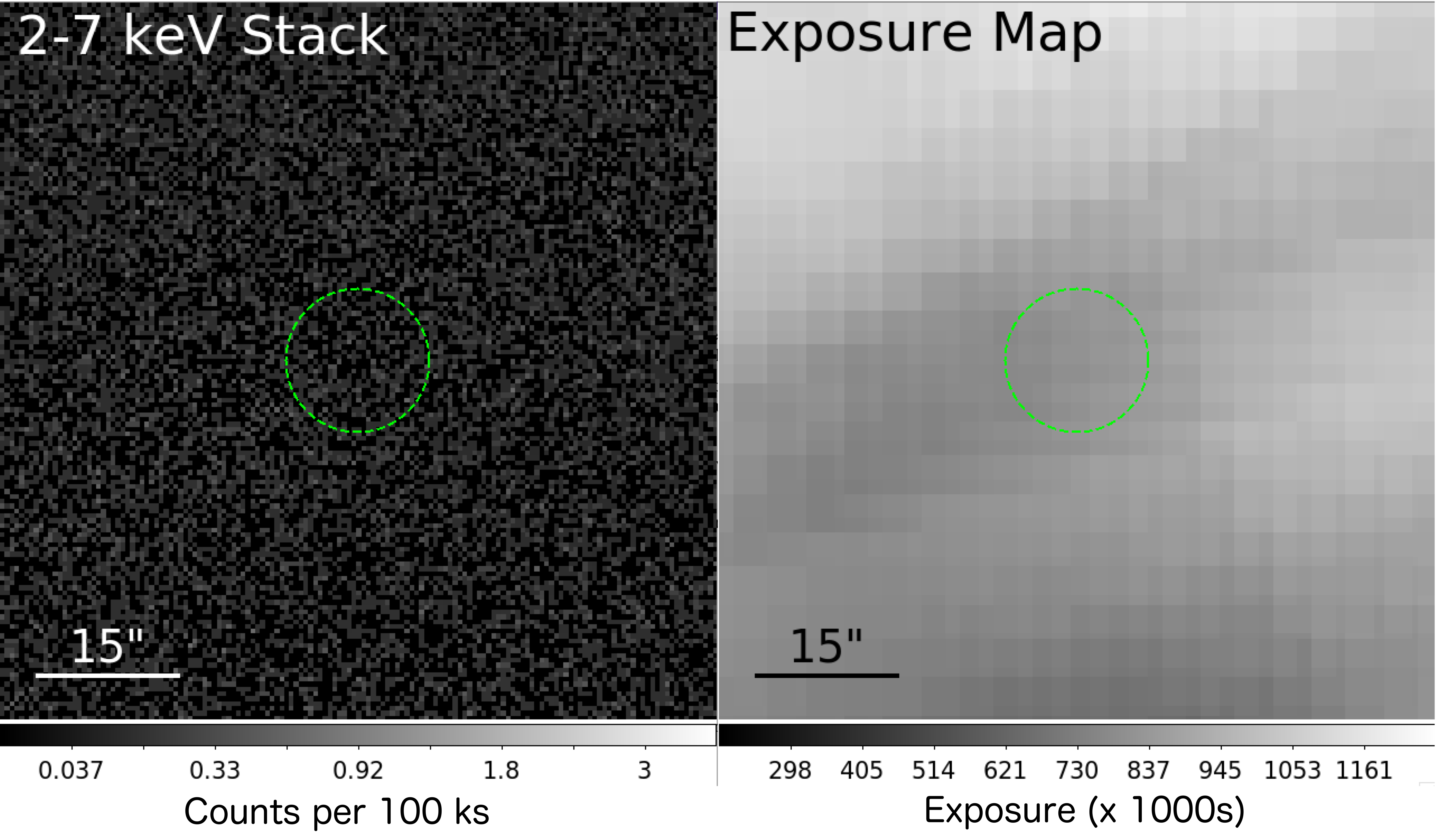}
 \vspace{-0.2cm}
 \caption{{Stacked 2-7~keV \textit{Chandra} image for 12 DF UDGs in the Coma cluster (in units of counts per 100~ks, left) and the exposure map in units of 1000~s (right). The galaxies and observations included are listed in Table~\ref{tab:datasources}. There is no detection within the 15~arcsec aperture shown in green, and the 3$\sigma$ upper limit (reported in the 0.3-10~keV bandpass) is $L_X < 2\times 10^{39}$~erg~s$^{-1}$.}}
 \label{fig:chandrastack}
\end{figure}

The \textit{Chandra} upper limits range from 0.3-10~keV $L_X < 1-7\times 10^{40}$~erg~s$^{-1}$ for the 12 DF galaxies in the field of view. When stacking the data {(Figure~\ref{fig:chandrastack})}, we obtain a limit of $L_X < 2\times 10^{39}$~erg~s$^{-1}$ for the average galaxy, with the caveat that the sample size is small and the detectors are different. {In addition, the exposure times are very different.} Hence, we prefer the individual limits. This is a worse constraint than the recent \citet{kovacs2019} result for isolated UDG candidates as they stacked more galaxies. Assuming that their galaxies are at an average distance of 65~Mpc they report an average $L_X < 3.1\times 10^{37}$~erg~s$^{-1}$. Another reason for their higher sensitivity is that we are limited by the cluster background.

Recently, \citet{kovacs2020} extended that analysis to Coma itself with an investigation of a much larger sample of UDGs from \citet{yagi2016}. They find at most two off-nuclear X-ray sources among Coma UDGs, which is fully consistent with our analysis that used much of the same data but restricted analysis to the DF sample.

The individual upper limits rule out any strong AGN activity. Considering that UDGs tend to have stellar masses consistent with dwarf galaxies ($M_* < 10^9 M_{\odot}$), and thus would be expected to host smaller central black holes if they follow a similar relation to high surface brightness galaxies, we may not be sensitive to weak Seyfert-like accretion. The sample size is too small to use the AGN frequency to test whether UDGs obey scaling relations measured in other systems, but if future surveys find few or no AGNs at luminosities near $10^{40}$~erg~s$^{-1}$, then it would indicate that any massive black holes in UDGs are more likely to scale with the stellar population rather than the (potentially much larger) dynamical mass. This is because the same luminosity probes lower Eddington ratios in larger black holes, as described in \citet{miller2012}.

These limits also far exceed the contributions from bright LMXBs, except perhaps the brightest ultraluminous X-ray sources (ULXs). ULXs are often associated with star formation, and so we do not expect to see any in the Coma UDGs. In the next section, we discuss the constraints on LMXBs in DF44 in more detail.

\section{Dragonfly 44}

\begin{figure}
 \includegraphics[width=\linewidth]{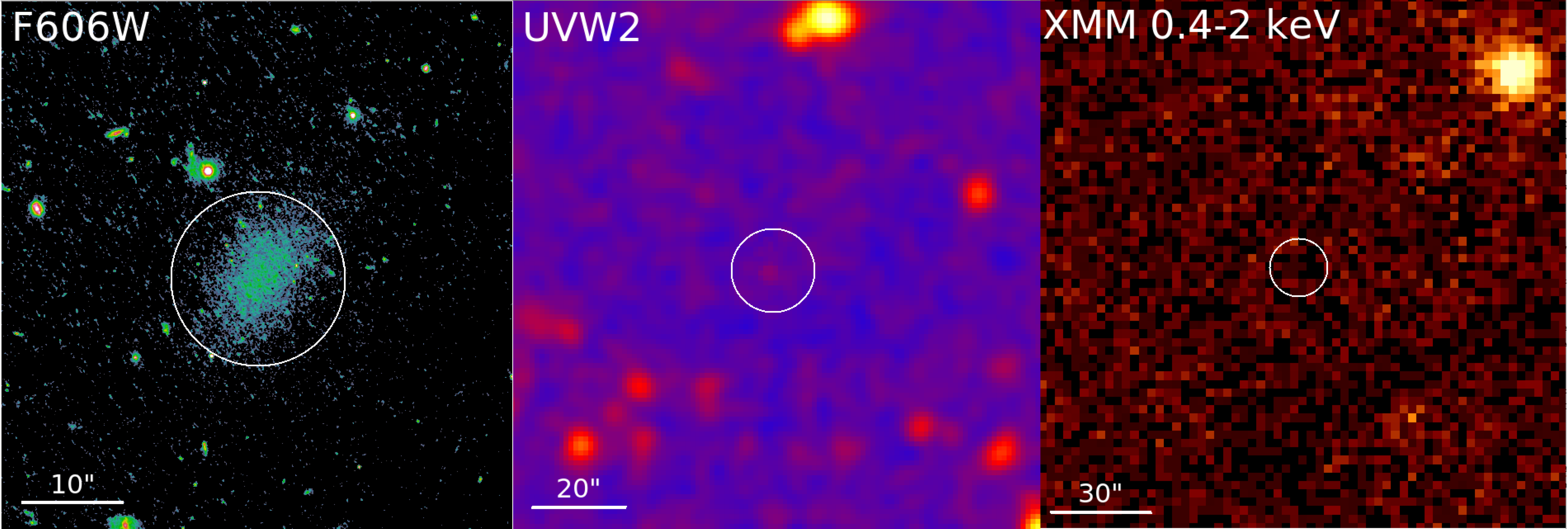}
 \caption{Optical (HST F606W), UV (\swift UVW2), and X-ray (\xmm 0.4-2 keV) images of DF44. The weak detection in the UV indicates little ongoing star formation, while the non-detection in the X-rays at $10^{38}$~erg~s$^{-1}$ indicates a lack of hot gas and low XRB activity. The circle is the same scale in each panel, and background was measured in annular regions outside the source region.}
 \label{fig:DF44_images}
\end{figure}

DF44 merits special consideration because it has some of the strongest evidence for a large dynamical mass and the deepest data {(Figure~\ref{fig:DF44_images})}. When using an elliptical aperture based on the \textit{HST} optical image (as opposed to a circular aperture based on $r_{\text{eff}}$, in which it was not detected), DF44 was detected in the NUV channel in a shallow archival GALEX exposure, {but it is detected in a circular aperture for both the \textit{Swift} and SDSS bands}. We used the UV data to calculate the SFR based on the relation from \cite{rozagonzalez2002}:
\begin{equation*}
\textrm{SFR}(UV)[\textrm{M}_{\odot}\textrm{yr}^{-1}] = 1.4\times10^{-28}L_{\nu}[\textrm{erg}~\textrm{s}^{-1} \textrm{Hz}^{-1}]
\end{equation*}
which is applicable within 1500-2400\AA\ where the considered integrated spectrum is nearly flat in star-forming regions (Table~\ref{tab:df44phot}). The deep UVW2 data (Figure~\ref{fig:DF44_images}) provide the strongest constraint, with SFR $=6\times 10^{-4} M_{\odot}$~yr$^{-1}$. This is significantly below the value for UVW1 (after correcting for the red leak) and NUV values of SFR~$\approx 2\times 10^{-3}$. As we noted above, the NUV light may not trace very young stars, and the difference between the NUV and UVW2 indicates that the NUV light traces stars formed in the past Gyr but not in the past few Myr. Indeed, even some of the UVW2 light likely comes from these older stars, so this SFR is an upper bound. 

Nevertheless, if we take SFR $=6\times 10^{-4} M_{\odot}$~yr$^{-1}$ as the steady-state value, in 1~Gyr DF44 would produce $6\times 10^5 M_{\odot}$ of new stars, which is only 0.2\% of its stellar mass ($3\times 10^8 M_{\odot})$. The NUV$-r$ color places DF44 (Figure~\ref{fig:dfcmrcomparisons}) firmly in the regime where there has been star formation in the past Gyr comprising $\sim$1\% of the stellar mass \citep{kaviraj2007}, suggesting that the SFR was higher in the past. Notably, DF44 is photometrically indistinct from the general DF population. This situation is consistent with the sample's NUV$-r$ colors and the finding by \cite{koda2015} that the Coma UDGs are not currently forming stars (based on H$\alpha$ luminosities). 

\begin{table*}
\caption{Optical and ultraviolet photometric information for DF44.\label{tab:df44phot}}
\begin{tabular*}{0.7\textwidth}{@{\extracolsep{\fill}}lccccc}
\hline
 {Filter} & {$\lambda_{\text{eff}}$} & {$F_{\lambda}(\times10^{-18})$}  & {$t_{\text{exp}}$}  & {Conv. Factor} & {SFR} \\
&  {[\AA]} &  {[erg~s$^{-1}$~cm$^{-2}$~\AA$^{-1}$]} &  {[s]} & &  {[$M_{\odot}$~yr$^{-1}$]}\\
\hline
 FUV             & 1516 & $<14$       & 1488  & (1) $1.40\times10^{-15}$  & \cds       \\
 UVW2            & 1928 & $2.8\pm1.3$ & 40876 & (1) $5.98\times10^{-16}$  & $6\times10^{-4}$\\
 NUV             & 2267 & $6.8\pm2.6$ & 1639  & (1) $2.06\times10^{-16}$  & $2\times10^{-3}$\\
 UVW1            & 2505 & $7.2\pm1.4$ & 30134 & (1) $4.21\times10^{-16}$  & $3\times10^{-3}$\\
 SDSS \textit{u} & 3580 & $37\pm4.1$  & 54.0  & (2) $8.63\times10^{-18}$  & \cds \\
 SDSS \textit{g} & 4754 & $42\pm3.6$  & 54.0  & (2) $4.96\times10^{-18}$  & \cds \\
 SDSS \textit{r} & 6166 & $42\pm3.3$  & 54.0  & (2) $2.86\times10^{-18}$  & \cds \\
\hline
\multicolumn{6}{p{0.6\textwidth}}{Conversion factors from standard calibration: (1) counts~s$^{-1}$ to $F_{\lambda}$ ; (2) nMgy to $F_{\lambda}$.}
\end{tabular*}
\end{table*}

Compared to the \textit{Chandra} upper limits on the X-ray luminosity for Coma DF galaxies, DF44 has a much tighter {0.3-10~keV, 3$\sigma$ upper limit} of $L_X< 1.0\times 10^{38}$~erg~s$^{-1}$, {although it is individually undetected (E. Miller 2019, private communication)}. This is only a few times higher than the \cite{kovacs2019} measurement from a stack of isolated UDG candidates, and it is sensitive enough to constrain the X-ray binary population, low level nuclear activity, and the presence of hot gas. In particular, it rules out the presence of any nuclear activity that could suppress star formation. This is not surprising, since most local early-type galaxies with similar stellar mass are not detected at this sensitivity \citep{gallo2010,miller2012}. However, we cannot rule out very low level or highly radiatively inefficient accretion, since Sgr~A* has a quiescent luminosity of a few $10^{33}$~erg~s$^{-1}$ \citep{baganoff2003}. 

The brightest non-ULX LMXBs have {$0.3-10$~keV} luminosities exceeding $10^{38}$~erg~s$^{-1}$, so it is worth examining whether the X-ray non-detection for DF44 constrains its LMXB population. The total LMXB luminosity is known to scale with the stellar mass \citep{gilfanov2004}, and using the \cite{lehmer2010} scaling of $L_{X,\text{LMXB}} = 9\times 10^{28} \times M_*$, the average LMXB luminosity for a DF44-sized galaxy is $3\times 10^{37}$~erg~s$^{-1}$, which is consistent with the upper limit from XMM. However, LMXBs are Poisson distributed according to the X-ray luminosity function \citep{gilfanov2004}, which is a power law in the regime of $L_X \sim 10^{38}$~erg~s$^{-1}$. Thus, there is a finite chance of detecting LMXBs above a given luminosity for any galaxy. To estimate this probability, we adopt the formalism of \cite{foord2017} and \cite{lee2019}, which uses the \cite{lehmer2010} scaling and luminosity function to estimate the average \textit{number} of LMXBs and treats it as the mean of a Poisson distribution. Then, galaxies are simulated drawing from this distribution and randomly assigning a luminosity weighted by the luminosity function. With a large number of draws from the sample, one can calculate the likelihood of detecting a total LMXB luminosity above some threshold. For DF44 and the $10^{38}$~erg~s$^{-1}$ limit, we find a 95\% chance of no detection. This amounts to a stronger statement that the non-detection of LMXBs is expected from the stellar mass. 

However, it is not clear whether LMXBs scale with stellar mass or GCs, since a large fraction of LMXBs may form in GCs before migrating into the rest of the galaxy. In the Milky Way galaxy, GCs host about 10\%  of the active LMXB population, while at the same time accounting for less than 1\%\ of the light, implying that LMXB formation in GCs must be hundreds of times more efficient than in the field \citep{clark1975}. This has long been attributed to enhanced dynamical formation mechanisms in the dense cores of GCs \citep{pooley2003}. Since the number of GCs and stellar mass both typically scale with dynamical mass, for most galaxies one cannot tell the difference. However, one piece of evidence in favor of a large dynamical mass in DF44 is its large number ($74\pm14$) of GCs identified in deep Gemini images \citep{vdokkum2016,vdokkum2017}. This number is consistent with a Milky Way-like dynamical mass, and much higher than expected for the stellar mass. 

For nearby galaxies, the odds of detecting a GC LMXB above $10^{38}$~erg~s$^{-1}$ are $\sim$5-10\%, with a mild dependence on morphological type. More specifically, the number of GC LMXBs brighter than $10^{38}$ erg~s$^{-1}$ is estimated by \cite{sarazin03} to be $1.5\times 10^{-7}$ per optical luminosity of GCs (or a 1-2\% chance detection for a $10^5$ solar mass GC). {While the actual number likely depends on other GC properties as well (such as color and concentration), overall, the percentage of GCs containing a bright LMXB ranges between a few up to 10\% \citep{sivakoff07}}.
Additionally, \cite{peacock2016} find that the XLF of GC XRBs is flatter than in the field, implying that the chances of detecting a bright LMXB are somewhat higher in GCs than they are in the field (this is not necessarily true at lower X-ray luminosities, where the XLF slope(/s) is(/are) flatter and the chance of detection depends more strongly on the actual slope value). In DF44, if we adopt a conservative value of a 5\% chance detection per GC and consider a total of 74 GCs, then we infer a 2\% probability of detecting \textit{no} GC LMXB in our data. The probability drops to 0.04\% if we assume a 10\% detection chance for a single GC. Overall, these numbers indicate that the GCs in DF44 may be fairly inefficient at producing high luminosity LMXBs. However, as shown by \cite{kundu2007}, GC metallicity is likely to play a major role in these estimates (with metal-rich GCs being three times as likely to host LMXBs). 

Finally, we estimated the sensitivity to a hot gas halo with $kT=0.2$~keV and $Z=0.3 Z_{\odot}$, which is based on the parameters measured around the Milky Way \citep{matt_miller15}. We used an elliptical aperture with semi-major and semi-minor axes $a=9^{\prime\prime}$ and $b=6^{\prime\prime}$, respectively, and a position angle matched to DF44 based on the HST image. The XMM non-detection limits the emission measure to EM $< 2.5\times 10^{-7}$~cm$^{-3}$ for these parameters, which implies a mass $M_{\text{hot}} < 6\times 10^6 M_{\odot}$ in the region, assuming an ellipsoidal, axisymmetric volume with $b=c=6^{\prime\prime}$ at $d\approx 100$~Mpc. This rules out a significant hot halo: assuming that there is this much gas and that it is distributed following the same $\beta$ model measured around the Milky Way \citep{matt_miller15}, with $\beta \approx 0.5$ and a core radius of 3-4~kpc, the limit to the mass within the virial radius is $M_{\text{hot}}(r<r_{\text{vir}}) < 10^8 M_{\odot}$. It is more likely that there is no extended hot halo, and that any hot gas in the region is essentially contiguous with the intracluster medium. 

\section{Discussion}

The UV and X-ray properties of DF44 and other Coma DF UDGs offer clues to their formation. The NUV$-r$ colors (Figure~\ref{fig:dfcmrcomparisons}) suggest that at least 1\% of the stellar mass has formed in the past Gyr \citep{kaviraj2007}, but their current SFR is much smaller (near $10^{-7} M_{\odot}$~yr$^{-1}$ on average and less than $6\times 10^{-4} M_{\odot}$~yr$^{-1}$ in DF44). This would lead to a buildup of less than $<$~0.2\% in stellar mass over 1~Gyr, which is consistent with \citet{koda2015}. It is possible that the NUV does not trace young stars, but instead hot horizontal branch stars \citep{OConnell1999}, which may be responsible for the ``UV upturn'' seen in elliptical galaxies \citep{Yi2011}. However, the sample-averaged FUV$-$NUV magnitude of 1.15 is inconsistent with the UV upturn \citep{Yi2011} and the colors of most of the detected UDGs are too blue (NUV$-r \lesssim 4$~mag) to explain apart from residual star formation. 

The discrepancy between ongoing and recent star formation indicates that the galaxies were recently quenched within the past 1~Gyr. One possible explanation is ram-pressure stripping \citep{jiang2019}, {which is based in part on their report of a $B-R$ color gradient. If so, we would also expect to see a color gradient in NUV$-r$, since galaxies that have recently passed through the Coma core should be quenched. Meanwhile, assuming that UDGs have a similar velocity dispersion to the brighter Coma members, it is possible that galaxies at the outskirts have yet to pass through the core.} For example, DF44 is offset from the Coma systemic velocity by 900~km~s$^{-1}$ and lies at a projected distance of 1.8~Mpc from the Coma core, placing a lower limit of 3~Gyr on its last possible core passage.

We find a weak gradient in NUV$-r$, but only because the average $r$ band magnitudes increase by $\approx$~0.5~mag from 0 to 2.5~Mpc, corresponding to a decrease in stellar mass of about a factor of two. There is no significant correlation between NUV flux and distance from the core. One reason may be that estimating the distance from the core by projected radius can be unreliable, and another is that, depending on when the UDGs formed, some fraction of UDGs at the outskirts have already passed through the core. We also do not find a significant correlation between $g$ or $i$ and distance from the core, but this is because of the considerably higher number of undetected galaxies in these bands. Regardless, the $r$ band correlation with projected radius provides tentative support to the \citet{jiang2019} scenario, which in turn explains our more robust result that complete quenching occurred in the past Gyr.

If UDGs have dark-matter fractions exceeding 99\%, then multiple Coma UDGs should be massive enough to retain hot halos from their formation \cite{white&frenk1991}. DF44 is sufficiently far from the cluster core (1.8~Mpc in projection) that it could have a hot corona if it did not pass through the denser regions of the intracluster medium. The lack of such a halo may indicate that it was previously stripped (hot gas is easier to strip than cold gas), which is consistent with the UV data, that it is not dynamically massive \citep{kovacs2019}, or that the hot halo failed to form. In any case, the absence of a halo implies that there is no hidden reservoir of material that could fuel future star formation, and the current lack of star formation and low stellar mass imply that no X-ray corona will be built up through stellar feedback. Meanwhile, using a larger sample, \citet{kovacs2020} argue that the lack of X-ray point sources in Coma UDGs implies that they are dwarf galaxies, which is consistent with our stacked X-ray upper limits and the expected number of X-ray binaries from the stellar mass (as opposed to GCs).

The only galaxy with relatively tight limits on LMXB emission is DF44, and the limit of $10^{38}$~erg~s$^{-1}$ is fully consistent with the expectation from its low stellar mass. However, if most LMXBs are formed in GCs, the lack of detection in DF44 is surprising. This suggests that DF44 GCs were inefficient at producing LMXBs, but deeper data would be needed to fully explore this issue. Future surveys with high-resolution X-ray optics could determine whether UDG GCs are generally inefficient at forming LMXBs, as multiple Coma UDGs have many more GCs than expected for their stellar masses. 

\section{Summary}

We investigated DF UDGs in the Coma cluster using new and archival X-ray and UV data. Our principal findings include:
\begin{enumerate}
    \item We clearly detect the UDGs in the FUV and NUV bands when stacking all the data. {7.5\% of galaxies (3/40) are individually detected in the FUV, and about 15\% (6/39) in the NUV}, although we note that the fields do not have uniform sensitivity in either band. The NUV$-r$ colors imply that at least 1\% of the stellar mass has formed within the past Gyr, but the average FUV magnitude implies that the SFR has since fallen by several orders of magnitude.
    \item There is a weak correlation between $r$-band magnitude and distance from the cluster core that implies that the stellar mass decreases by a factor of $\sim$2 from 0 to 2.5~Mpc. On the other hand, no correlation is seen in the NUV. Together, these imply that recent star formation has been slightly more efficient in the cluster outskirts, but with galaxy-to-galaxy scatter.
    \item There is no significant AGN activity in the 13 galaxies for which there are X-ray data. DF44 also has no bright X-ray binaries, which is expected if LMXBs track stellar mass but not if they track GCs, of which DF44 has many. The lack of X-ray detections is similar to what is found in isolated UDGs \citep{kovacs2019} and consistent with a wider study of Coma UDGs \citep{kovacs2020}. 
    \item DF44 has little or no hot gas bound to its potential ($M_{\text{hot}} < 6\times 10^6 M_{\odot}$ within the XMM aperture, or $<10^8 M_{\odot}$ within the virial radius), whereas one would expect a hot halo for a dynamical mass near $10^{12} M_{\odot}$.
\end{enumerate}
Overall, the drop in SFR and the correlation of $r$ magnitude with projected distance from the Coma core is consistent with quenching by ram-pressure stripping, as suggested by \citet{jiang2019}. Ram-pressure stripping would also explain the loss of the hot gaseous halo even if DF44 is dynamically massive. However, deeper UV data are required to confirm this hypothesis, as any correlation between NUV flux and projected distance is too weak to find in existing data.

\section{Acknowledgments}

The authors thank the reviewer for helpful comments and suggestions that improved the manuscript. C.~L. and E.~G. gratefully acknowledge support for this work under the NASA award \#80NSSC18K0376. 

This work is based on observations obtained with \textit{XMM-Newton}, an ESA science mission with instruments and contributions directly funded by ESA Member States and NASA.

This research has made use of data and/or software provided by the High Energy Astrophysics Science Archive Research Center (HEASARC), which is a service of the Astrophysics Science Division at NASA/GSFC and the High Energy Astrophysics Division of the Smithsonian Astrophysical Observatory.

Funding for SDSS-III has been provided by the Alfred P. Sloan Foundation, the Participating Institutions, the National Science Foundation, and the U.S. Department of Energy Office of Science. The SDSS-III web site is http://www.sdss3.org/.

SDSS-III is managed by the Astrophysical Research Consortium for the Participating Institutions of the SDSS-III Collaboration including the University of Arizona, the Brazilian Participation Group, Brookhaven National Laboratory, Carnegie Mellon University, University of Florida, the French Participation Group, the German Participation Group, Harvard University, the Instituto de Astrof\'isica de Canarias, the Michigan State/Notre Dame/JINA Participation Group, Johns Hopkins University, Lawrence Berkeley National Laboratory, Max Planck Institute for Astrophysics, Max Planck Institute for Extraterrestrial Physics, New Mexico State University, New York University, Ohio State University, Pennsylvania State University, University of Portsmouth, Princeton University, the Spanish Participation Group, University of Tokyo, University of Utah, Vanderbilt University, University of Virginia, University of Washington, and Yale University.

\bibliographystyle{mnras}

\end{document}